\documentclass[onecolumn]{emulateapj}
\usepackage{apjfonts}
\usepackage{amsbsy}
\usepackage{graphicx}
\epsfclipon

\newcommand{\etal}{{et al.\hskip 3pt}}

\newcommand{\eg}{{e.g.,\hskip 3pt}}

\def\4he{$^4$He}
\def\3he{$^3$He}
\def\7li{$^7$Li}
\def\6li{$^6$Li}
\def\Yp{Y$_{\rm P}$~}
\def\yd{$y_{\rm D}$~}

\def\hii{H\thinspace{$\scriptstyle{\rm II}$}~}

\def\heii{He\thinspace{$\scriptstyle{\rm II}$}~}
\def\Nnu{N$_{\nu}$~}
\def\eten{$\eta_{10}$~}
\newcommand{\epm}{\ensuremath{e^{\pm}\;}}

\newcommand{\be}{\begin{equation}}
\newcommand{\ee}{\end{equation}}

\begin{document}


\title{Constraining The Early-Universe Baryon Density And Expansion Rate}

\author{Vimal Simha\altaffilmark{1} and Gary Steigman\altaffilmark{2}}   

\altaffiltext{1}{Department of Astronomy, The Ohio State University, 140 
West 18th Ave., Columbus, OH 43210}
\altaffiltext{2}{Departments of Physics and Astronomy and Center for Cosmology 
and Astro-Particle Physics, The Ohio State University, 191 West Woodruff 
Ave., Columbus, OH 43210}


\begin{abstract}

We explore the constraints on those extensions to the standard models 
of cosmology and particle physics which modify the early-Universe, 
radiation-dominated, expansion rate $S \equiv H'/H$ (parametrized by 
the effective number of neutrinos N$_{\nu}$).  The constraints on $S$ 
(N$_{\nu}$) and the baryon density parameter $\eta_{\rm B} \equiv (n_{\rm B}/n_{\gamma}) = 10^{-10}\eta_{10}$, derived from Big Bang Nucleosynthesis 
(BBN, $t \sim 20$ minutes) are compared with those inferred from the 
Cosmic Microwave Background Anisotropy spectrum (CMB, $t \sim 400$~kyr) 
and Large Scale Structure (LSS, $t \sim 14$~Gyr).  At present, BBN 
provides the strongest constraint on \Nnu (N$_{\nu} = 2.4\pm0.4$ at 
68\% confidence), but a weaker constraint on the baryon density.  In 
contrast, while the CMB/LSS data best constrain the baryon density 
($\eta_{10} = 6.1^{+0.2}_{-0.1}$ at 68\% confidence), independent of 
N$_{\nu}$, at present they provide a relatively weak constraint on 
\Nnu which is, however, consistent with the standard value of \Nnu = 
3.  When the best fit values and the allowed ranges of these 
CMB/LSS-derived parameters are used to calculate the BBN-predicted 
primordial abundances, there is excellent agreement with the 
observationally inferred abundance of deuterium and good agreement 
with \4he, confirming the consistency between the BBN and CMB/LSS
results.  However, the BBN-predicted abundance of \7li is high, by 
a factor of 3 or more, if its observed value is uncorrected for 
possible dilution, depletion, or gravitational settling.  We 
comment on the relation between the value of \Nnu and a possible 
anomaly in the matter power spectrum inferred from observations 
of the Ly-$\alpha$ forest.  Comparing our BBN and CMB/LSS results
permits us to constrain any post-BBN entropy production as well 
as the production of any non-thermalized relativistic particles.
The good agreement between our BBN and CMB/LSS results for \Nnu 
and $\eta_{\rm B}$ permits us to combine our constraints finding, 
at 95\% confidence, 1.8 $<$ \Nnu $<$ 3.2 and 5.9 $<$ \eten $<$ 6.4.

\end{abstract} 

\section{Introduction}

The standard models of particle physics and of cosmology with dark 
energy, baryonic matter, radiation (including three species of light 
neutrinos), and dark matter is consistent with cosmological data from 
several widely-separated epochs.  However, there is room to accommodate 
some models of non-standard physics within the context of this well 
tested, $\Lambda$CDM cosmology.  One possibility, explored here, is 
that of a non-standard expansion rate ($S \equiv H'/H$, where $H$ is 
the Hubble parameter) during the early, radiation-dominated evolution 
of the Universe driven perhaps, but not necessarily, by a non-standard 
content of relativistic particles ($\rho_{\rm R}' \neq \rho_{\rm R}$), 
parametrized by the equivalent number of additional neutrinos 
($\rho_{\rm R}' \equiv \rho_{\rm R} + \Delta$N$_{\nu}\rho_{\nu}$, 
where $\Delta$N$_{\nu} \equiv$~N$_{\nu} - 3$ prior to the epoch of 
\epm annihilation).  In the epoch just prior to \epm annihilation, 
which is best probed by Big Bang Nucleosynthesis (BBN), $\rho_{\rm R} 
= \rho_{\gamma} + \rho_{e} + 3\rho_{\nu} = 43\rho_{\gamma}/8$, so that
\be
S^{2} \equiv \left({H' \over H}\right)^{2} \equiv {\rho_{\rm R}' \over \rho_{\rm R}} 
\equiv 1 + {7\Delta{\rm N}_{\nu} \over 43}.
\ee 
In the epoch after the completion of \epm annihilation, best probed 
by the Cosmic Microwave Background (CMB) and by Large Scale Structure 
(LSS), the relations between $\rho_{\rm R}$ and $\rho_{\gamma}$ and 
between $S$ and $\Delta$N$_{\nu}$ differ from those in eq.~1, as 
described below in some detail in \S2.

We use the comparison between the predicted and observed 
abundances of the light elements produced during BBN, and between 
the predicted and observed CMB anisotropy spectrum, along with data 
from LSS observed in the present/recent Universe, to constrain new 
physics which leads to a non-standard, early-Universe expansion rate 
($S$) or, equivalently, to place bounds on the effective number of 
neutrinos (N$_{\nu}$); for related earlier work see, \eg \cite{barger03,cyburt04}.  In addition, 
the baryon density and, any variation in it over widely-separated 
epochs in the evolution of the Universe, are constrained simultaneously 
with N$_{\nu}$, thereby testing the standard-model expectation that 
the ratio (by number) of baryons to CMB photons ($\eta_{\rm B} \equiv 
n_{\rm B}/n_{\gamma}$) should be unchanged from \epm annihilation ($T 
\la 1/2$ MeV; $t \ga 3$~s) until the present ($T = 2.725$~K $\approx 
2\times 10^{-10}$~MeV, $t \approx 14$~Gyr).  

In \S2 a non-standard expansion rate ($S$) is related to a non-standard 
radiation density, parametrized by an effective number of neutrinos 
N$_{\nu}$.  In \S3, the observationally inferred primordial light element 
(D, \4he, \7li) abundances are used to constrain the radiation density 
(expansion rate) and the baryon density.  In \S4, we assume a flat, 
$\Lambda$CDM cosmology and use data from the CMB, along with a prior 
on the Hubble parameter from the HST key project, luminosity distances 
of type Ia supernovae, and the matter power spectrum to provide 
independent constraints on the radiation density and the baryon 
density.  The results from these two epochs, very widely-separated 
in time, are compared constraining any post-BBN entropy production.
The good agreement between them permits us to combine them to obtain 
a joint constraint on the baryon density parameter ($\eta_{\rm B}$) 
and the effective number of neutrinos (N$_{\nu}$).

\section{Non-Standard Early-Universe Expansion Rate or Radiation Density}

In the radiation-dominated early universe ($\rho_{\rm TOT} \rightarrow 
\rho_{\rm R}$) the expansion rate ($H$) is related to the radiation 
density through the Friedman equation.  
\be
H^{2} = {8\pi \over 3}G\rho_{\rm R}.
\ee
Any modification to the radiation density, or to the Friedman equation 
by a term which evolves like the radiation density (as the inverse 
fourth power of the scale factor) can be parametrized by an equivalent 
number of additional neutrinos $\Delta$N$_{\nu}$ where, prior to \epm 
annihilation, $\Delta$N$_{\nu} \equiv~$N$_{\nu} - 3$.  For the standard 
models of particle physics and cosmology, in the epoch after muon 
annihilation ($T \la 100$~MeV) and prior to $e^{\pm}$ annihilation 
($T \ga 0.5$~MeV), the radiation consists of an equilibrium mixture 
of photons, relativistic \epm pairs, and three flavors of extremely 
relativistic, left-handed neutrinos (and their right-handed, 
antineutrino counterparts).  In this case, the total radiation 
density may be written in terms of the photon density as
\be
\rho_{\rm R} = {43 \over 8}\rho_{\gamma} = 5.375\rho_{\gamma}.
\ee  
In this same epoch, prior to \epm annihilation, a modified radiation 
density can be written as
\be
\rho'_{\rm R} = \rho_{\rm R} \left(1+\frac{7 \Delta {\rm N}_{\nu}}{43}\right) = 
\rho_{\rm R} (1 + 0.163\Delta {\rm N}_{\nu})
\ee
where $\rho_{\rm R}$ is the standard-model radiation energy density and 
$\rho'_{\rm R}$ is the modified, non-standard model radiation energy 
density.  In a sense, this modified energy density is simply a proxy 
for a non-standard expansion rate during the radiation-dominated epoch
relevant for comparison with BBN.  
\be
S \equiv \frac{H'}{H} = \left(\frac{\rho'_{\rm R}}{\rho_{\rm R}}\right)^{1/2} = 
\left(1+\frac{7 \Delta {\rm N}_{\nu}}{43}\right)^{1/2}.
\ee
After \epm annihilation the surviving relativistic particles are the
photons (which will redshift to the presently observed CMB) and the
now decoupled, relic neutrinos.  In the approximation that the neutrinos 
are fully decoupled at \epm annihilation, the post-\epm annihilation 
photons are hotter than the neutrinos by a factor of $T_{\gamma}/T_{\nu} 
= (11/4)^{1/3}$ and
\be
\rho_{\rm R} = \left[1 + 3\times {7 \over 8}\left({4 \over 11}\right)^{4/3}\right]\rho_{\gamma} 
= (1 + 3\times 0.227)\rho_{\gamma} = 1.681\rho_{\gamma},
\ee
so that
\be
\rho'_{\rm R} = \rho_{\gamma}(1.681 + 0.227 \Delta {\rm N}_{\nu}) = 
\rho_{\rm R}(1 + 0.135 \Delta {\rm N}_{\nu}),
\ee
where, as before, $\Delta$N$_{\nu} \equiv$~N$_{\nu} - 3$.

However, it is well known \citep{dicus, dt, hm, lopez, dolgov} that the 
standard-model neutrinos are not fully decoupled at \epm annihilation.  
As a result, the relic neutrinos share some of the energy/entropy 
released by \epm annihilation and they are warmer than in the fully 
decoupled approximation, increasing the ratio of the post-\epm 
annihilation radiation density to the photon energy density.  
While the post-\epm annihilation phase space distribution of the 
decoupled neutrinos is no longer that of a relativistic, Fermi-Dirac 
gas, according to~\cite{mangano05} the additional contribution to 
the total energy density can be accounted for by replacing \Nnu 
= 3 with \Nnu = 3.046, so that
\be
\rho_{\rm R} \rightarrow (1 + 3.046\times 0.227)\rho_{\gamma} =
1.692\rho_{\gamma}.
\ee
For deviations from the standard model that can be treated 
as equivalent to contributions from fully decoupled neutrinos,
\be
\rho'_{\rm R} = \rho_{\rm R}(1 + 0.134\Delta {\rm N}_{\nu}),
\ee
where $\Delta$N$_{\nu} \equiv$~N$'_{\nu} - 3.046$ in the post-\epm 
annihilation Universe relevant for comparison with the CMB and 
LSS.  Note that in the standard model, where N$_{\nu} = 3$ prior 
to \epm annihilation, the neutrino contribution to the post-\epm 
annihilation radiation energy density is equivalent to N$'_{\nu} 
= 3.046$, so that for the standard models of cosmology (standard 
expansion rate) and of particle physics (standard radiation energy 
density), \Nnu = 3, N$'_{\nu} = 3.046$ and, $\Delta$N$_{\nu} = 0$.

We emphasize that although the non-standard radiation density 
(expansion rate) has been parametrized as if it were due to 
additional species of neutrinos, this parametrization accounts 
for any term in the Friedman equation whose energy density varies 
as $a^{-4}$, where $a$ is the scale factor.  From this perspective, 
$\Delta$N$_{\nu}$ could either be positive or negative; the latter 
does not necessarily imply fewer than the standard-model number 
of neutrinos but could, for example, be a sign that the three 
standard-model neutrinos failed to be fully populated in the 
early Universe or, could reflect modifications to the 3 + 1 
dimensional Friedman equations arising from higher-dimensional 
extensions of the standard model of particle physics \citep{rs, 
binetruy, cline}.

\section{N$_{\nu}$ and $\eta_{\rm{B}}$ From BBN}

The stage is being set for BBN when the Universe is about a tenth 
of a second old and the temperature is a few MeV.  At this time the 
energy density of the universe is dominated by relativistic particles.  
When the temperature drops below $\sim 2$~MeV, the neutrinos begin 
to decouple from the photon-\epm plasma.  However, they do continue 
to interact with the neutrons and protons through the charged-current 
weak interactions ($n + \nu_{e} \leftrightarrow p + e^{-}, p + 
\bar{\nu}_{e} \leftrightarrow n + e^{+}$, $n \leftrightarrow p + 
e^{-} + \bar{\nu}_{e}$), maintaining the neutron-to-proton ratio at 
its equilibrium value of $n/p =~$exp$(-\Delta m/kT)$, where $\Delta 
m$ is the neutron-proton mass difference.  When the temperature drops 
below $\sim 0.8$~MeV, and the Universe is $\sim 1$ second old, the 
reactions regulating the neutron-proton ratio become slower than 
the universal expansion rate ($\Gamma_{wk} < H$).  As a result, the 
neutron-proton ratio deviates from (exceeds) its equilibrium value, so 
that $n/p >$~exp$(-\Delta m/kT)$, and the actual $n/p$ ratio depends on 
the competition between the expansion rate ($H$) and the charged-current 
weak interaction rate ($\Gamma_{wk}$), as well as on the neutron decay 
rate, $1/\tau_{n}$, where $\tau_{n}$ is the neutron lifetime. 

Although nuclear reactions such as $n + p \leftrightarrow {\rm D} + 
\gamma$, proceed rapidly during these epochs, the large $\gamma$-ray 
background (the blue-shifted CMB) ensures that the deuterium (D) 
abundance is very small, inhibiting the formation of more complex 
nuclei.  The more complex nuclei begin to form only when $T \la 0.08$~MeV, 
after $e^{\pm}$ annihilation, when the Universe is about 3 minutes old.  
At this time the number density of photons with sufficient energy to 
photodissociate deuterium is comparable to the baryon number density 
and various two-body nuclear reactions can begin to build more complex 
nuclei.  Note that the neutron-to-proton ratio has decreased slightly 
since ``freeze-out" (at $T \la 0.8$~MeV) through the residual two body 
reactions as well as via beta decay.  Once BBN begins neutrons and proton 
combine to form D, \3he, and \4he.  The absence of a stable mass-5 nuclide 
presents a road-block to the synthesis of heavier elements in the expanding, 
cooling Universe, ensuring that the abundances of heavier nuclides are 
severely depressed below those of the lighter nuclei.  In standard BBN 
(SBBN) only D, \3he, \4he, and \7li are produced in astrophysically 
interesting abundances (for a recent review see \citet{steigman07}).  
While the BBN-predicted abundances of D, \3he, and \7li are most 
sensitive to the baryon density, that of \4he is very sensitive 
to the neutron abundance when BBN begins and, therefore, to the 
competition between the weak-interaction rate and the universal 
expansion rate.  The primordial abundances of D, \3he, or \7li 
are baryometers, constraining $\eta_{\rm B}$, while  \4he mass 
fraction (Y$_{\rm P}$) is a chronometer, depending mainly on $S$ 
or, N$_{\nu}$.  In the standard model of particle physics and 
cosmology with three species of neutrinos and their respective 
antineutrinos, the primordial element abundances depend on only 
one free parameter, the baryon density parameter, the post-\epm 
ratio (by number) of baryons to photons, $\eta_{\rm B}$ = 
$n_{\rm B}/n_{\gamma}$.  This parameter may be related to 
$\Omega_{\rm B}$, the present-Universe ratio of the baryon mass 
density to the critical mass-energy density (see \citet{steigman06})
\be
\eta_{10} = 10^{10}~n_{\rm B}/n_{\gamma} = 273.9~\Omega_{\rm B}h^2.
\ee
The abundance of \4he is very sensitive to the early expansion rate.  
Since a non-standard expansion rate ($S \neq 1$) would result in fewer 
or more neutrons at BBN and, since most neutrons are incorporated into 
\4he, the predicted \4he abundance differs from that in SBBN ($S = 1$; 
$\Delta$N$_{\nu}$ = 0).  In contrast, the abundance of \4he is not very 
sensitive to the baryon density since, to first order, all the neutrons 
available at BBN are rapidly converted to \4he.  For $\eta_{10} \approx 
6$, \Nnu $\approx 3$ and, for a primordial \4he mass fraction in the range 
$0.23 \la$~Y$_{\rm P} \la 0.27$, to a very good approximation \citep{kneller04, 
steigman07},
\be
{\rm Y}_{\rm P} = 0.2485 \pm 0.0006 + 0.0016[(\eta_{10} - 6) + 100(S - 1)].
\label{y_p}
\ee
In eq.~\ref{y_p}, the effect of incomplete neutrino decoupling on the 
\4he mass fraction is accounted for according to~\cite{mangano05} and 
$S$ is related to $\Delta$N$_{\nu}$ ($\Delta$N$_{\nu} \equiv ~$N$_{\nu} 
- 3.0$) by eq.~5.  As a result, for a fixed \4he abundance, a variation 
in $\eta_{10}$ of $\sim \pm~0.2$ (corresponding to a $\sim 3$\% uncertainty 
in the baryon density) is equivalent to an uncertainty in $\Delta$N$_{\nu}$ 
of $\sim \pm~0.02$.

In contrast to \4he, since the primordial abundances of D, \3he 
and \7li are set by the competition between two body production and 
destruction rates, they are more sensitive to the baryon density 
than to the expansion rate.  For example, for  $\eta_{10} \approx 
6$ and for a primordial ratio of D to H by number, \yd $\equiv
10^{5}({\rm D}/{\rm H})_{\rm P}$, in the range, $2 \la y_{\rm D} 
\la 4$, to a very good approximation \citep{kneller04, steigman07},
\be
y_{\rm D} = 2.64(1 \pm 0.03)\left[{6 \over \eta_{10} - 6(S - 1)}\right]^{1.6}.
\ee
The effect of incomplete neutrino decoupling on this prediction is
at the $\sim 0.3$\% level \citep{mangano05}, about ten times smaller 
than the overall error estimate above.

\subsection{Observationally-Inferred Primordial Abundances}

Given the monotonic post-BBN evolution of deuterium (as gas is cycled 
through stars, deuterium is destroyed) and the significant dependence 
of its predicted BBN abundance on the baryon density ($y_{\rm D\rm P} 
\propto \eta_{10}^{-1.6}$), deuterium is the baryometer of choice among 
the light nuclides produced during primordial nucleosynthesis.  While 
observations of D/H in the solar system and the local interstellar medium 
provide a lower limit to the relic deuterium abundance, it is the D/H 
ratio (by number) measured from observations of high redshift, low 
metallicity QSO absorption line systems which provide an estimate of 
its primordial abundance.  The weighted mean of the six, high redshift, 
low metallicity D/H ratios from \citet{kirkman03} and \citet{omeara06} is 
\citep{steigman07}
\be
y_{\rm D\rm P}=2.68^{+0.27}_{-0.25}.
\ee

Since the post-BBN evolution of \3he is more complex and model
dependent than that of deuterium and, since \3he is only observed 
in chemically-evolved \hii regions in the Galaxy and, since the \3he 
primordial abundance is only weakly dependent on the baryon density 
($10^{5}(^{3}$He/H)~$\equiv y_{3\rm P} \propto \eta_{10}^{-0.6}$), 
its role as a baryometer is limited.  

As for deuterium, the post-BBN evolution of \4he is monotonic, with 
\Yp increasing along with increasing metallicity.  At low metallicity, 
the \4he abundance should approach its primordial value.  As a result, 
it is the observations of helium and hydrogen recombination lines from 
low-metallicity, extragalactic \hii regions which are most useful in 
determining Y$_{\rm P}$.  At present, corrections for systematic 
uncertainties (and their uncertainties) dominate estimates of the 
observationally-inferred \4he primordial mass fraction and, especially, 
of its error.  Following \citet{steigman07}, we adopt for our estimate here,
\be
{\rm Y}_{\rm P}=0.240\pm0.006.
\ee
While the central value of \Yp adopted here is low, the 
conservatively-estimated uncertainty is relatively large (some 
ten times larger than the uncertainty in the BBN-predicted 
abundance for a fixed baryon density).  In this context, it
should be noted that although very careful studies of the 
systematic errors in very limited samples of \hii regions 
provide poor estimators of \Yp as a result of their uncertain 
and/or model-dependent extrapolation to zero metallicity, 
they are of value in providing a robust {\it upper bound} to 
\Yp.  Using the results of \cite{os}, \cite{fk}, and \cite{peimbert07}, 
we follow \citet{steigman07} in adopting,
\be
{\rm Y}_{\rm P} \\< 0.251\pm 0.002.
\ee

Although the BBN-predicted \7li relic abundance provides a potentially 
sensitive baryometer ((Li/H)~$\propto \eta_{10}^{2}$, for $\eta_{10} 
\ga 3$), its post-BBN evolution is complicated and model-dependent.  
For these reasons, it is the observations of lithium in the oldest, 
most metal-poor stars in galactic globular clusters and in the halo 
of the Galaxy which have the potential to provide the best estimate 
of the primordial abundance of \7li.  The complication associated 
with this approach is that these oldest galactic stars have had 
the most time to dilute or deplete their lithium surface abundances, 
leading to the possibility that the observed abundances require large, 
uncertain, and model-dependent corrections in order to infer the 
primordial abundance of \7li.  In the absence of corrections for 
depletion, dilution, or gravitational settling, the data of 
\citet{ryan} and \citet{asplund06} suggest
\be
[{\rm Li}]_{\rm P} \equiv 12+{\rm log (Li/H)} = 2.1\pm0.1.
\ee
In contrast, in an attempt to correct for evolution of the surface 
lithium abundances, \citet{korn06} use their observations of a small, 
selected sample of stars in the globular cluster NGC6397, along with 
stellar evolution models which include the effect of gravitational 
settling to infer
\be
[{\rm Li}]_{\rm P} = 2.54\pm0.1.
\ee

In the following analysis, the inferred primordial abundances of D and 
\4he adopted here are used to estimate $\eta_{10}$ and $\Delta$N$_{\nu}$.  
Given the inferred best values and uncertainties in these two parameters, 
the corresponding BBN-predicted abundance of \7li can be derived and 
compared to its observationally inferred abundance.

\subsection{BBN Constraints On N$_{\nu}$ And $\eta_{\rm B}$}


\begin{figure}
\centerline{\epsfxsize=3truein\epsffile{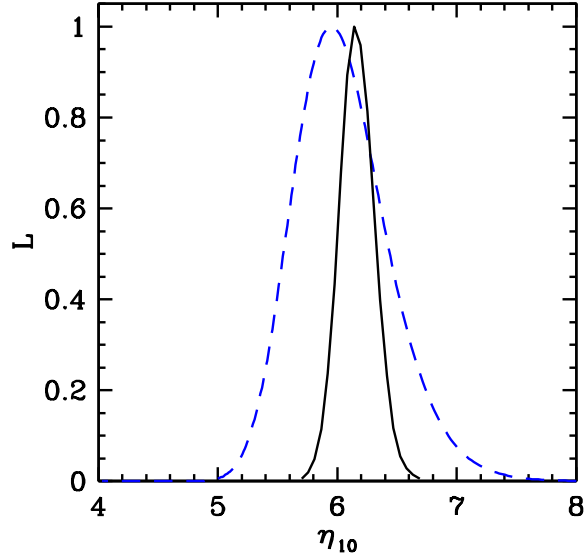}}
\caption{
The probability distribution of the baryon density parameter, 
$\eta_{10}$. The dashed line shows the probability distribution 
inferred from SBBN (N$_{\nu} = 3$) and the adopted primordial 
abundance of deuterium (see \S3).  The solid line is the 
probability distribution of $\eta_{10}$ inferred for N$_{\nu} 
= 3$ from the combination of the WMAP-5yr data, small scale 
CMB data, matter power spectrum data from 2dFGRS and SDSS LRG, 
SNIa, and the HST Key Project (see \S4).}
\label{fig:eta10sbbn}
\end{figure}


Since the primordial abundance of deuterium is most sensitive to 
$\eta_{10}$, while that of \4he is most sensitive to N$_{\nu}$, 
isoabundance contours of D/H and \Yp in the \{$\eta_{10}, 
\Delta$N$_{\nu}$\} plane are very nearly orthogonal; see 
\citet{kneller04}.  The analytic fits to BBN from \citet{kneller04}, 
updated by \citet{steigman07}, are used in concert with the primordial 
abundances of D and \4he adopted here to infer the best values, and 
to constrain the ranges of $\eta_{10}$ and $\Delta$N$_{\nu}$.  While 
these fits do have a limited range of applicability, they are, in 
fact, accurate within their quoted uncertainties for the range of 
parameter values and observed abundances considered here. 

In SBBN, with three species of neutrinos ($\Delta$N$_{\nu}=0$), the 
primordial abundances are only functions of the baryon density, 
$\eta_{\rm B}$.  For SBBN, the primordial deuterium abundance 
adopted in \S3.1, $y_{\rm D\rm P}=2.68^{+0.27}_{-0.25}$, implies,
\be
\eta_{10}({\rm SBBN})=6.0\pm0.4
\ee
This result is in excellent agreement with the independent estimate 
of $\eta_{10}=6.1^{+0.2}_{-0.1}$ from the CMB and LSS (discussed below 
in \S4).  The probability distributions of $\eta_{10}$ inferred from 
SBBN and from the CMB and LSS are shown in Figure \ref{fig:eta10sbbn}.

\begin{figure}
\centerline{\epsfxsize=3truein\epsffile{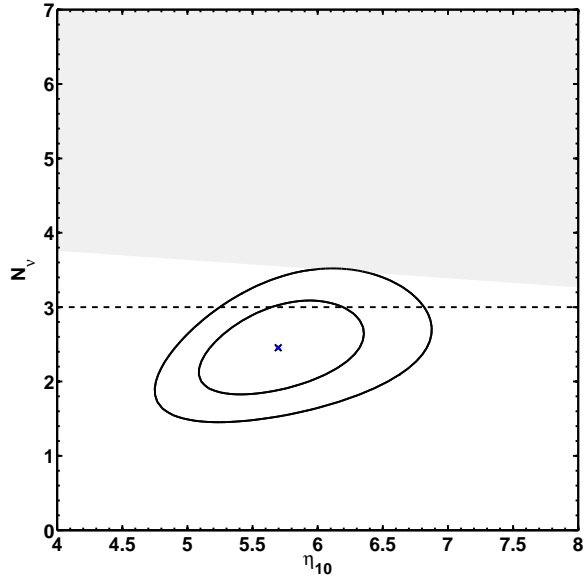}}
\caption {
The 68\% and 95\% contours in the \Nnu - $\eta_{10}$ plane 
derived from a comparison of the observationally-inferred 
and the BBN-predicted primordial abundances of D and \4he.
The shaded region is excluded by the 95\% upper bound to 
the helium abundance in eq.~15 (see, eq.~20).}
\label{fig:nnue10bbn}
\end{figure}

For non-standard BBN, with $\Delta$N$_{\nu} \neq 0$ ($S \neq 1$), there 
is a second free parameter, N$_{\nu}$ (or, $S$).  In this case, in 
addition to $y_{\rm D\rm P}$, the \4he abundance \Yp is used to constrain 
the \{$\eta_{10}$, N$_{\nu}$\} pair.  Adopting the D and \4he abundances 
from \S3.1 (Y$_{\rm P} = 0.240\pm0.006$), along with the analytic fits 
in eq. (11) and eq. (12), leads to, 
\be
\eta_{10}=5.7\pm0.4 \ , \ \ {\rm N}_{\nu}=2.4\pm0.4.
\ee
In Figure \ref{fig:nnue10bbn} are shown the 68\% and 95\% contours 
in the N$_{\nu} - \eta_{10}$ plane which follow from a comparison 
of the BBN predictions with the observationally-inferred primordial 
abundances of D and \4he.  Notice that while the best fit value 
of \Nnu is less than the standard-model value of \Nnu = 3, the 
standard-model value is consistent with the relic abundances at 
the $\sim 68$\% level.

These results are sensitive to the choices of the relic abundances
of D and \4he.  We note that if deuterium is ignored and the robust 
upper bound to the \4he mass fraction, Y$_{\rm P} < 0.255$ at 95\% 
confidence (eq.~15), is adopted, then eq.~11 provides an upper 
limit to $S$ (N$_{\nu}$) as a function of the baryon density,
\be
S < 1.10 - 0.01\eta_{10}.
\ee
For \eten in the range $5 \leq \eta_{10} \leq 7$ (see \S4), this 
corresponds to a robust upper bound to \Nnu ranging from 3.6 to 3.4. 

\section{N$_{\nu}$ And $\eta_{\rm B}$ From The CMB And LSS}

The pattern of temperature fluctuations in the cosmic microwave 
background contain information about the baryon density and the 
radiation density and thus serve as complementary probes of 
$\eta_{\rm B}$ and \Nnu some $\sim10^5$ years after BBN.  

The baryon density, parametrized by $\eta_{\rm B}$ or $\Omega_{\rm B}h^2$ 
affects the relative amplitudes of the peaks in the CMB temperature 
power spectrum.  The ratios of the amplitudes of the odd peaks to 
the even peaks provide a determination of $\eta_{\rm B}$ that is 
largely uncorrelated with N$_{\nu}$.  

The radiation density, parametrized by an effective number of neutrino 
species N$_{\nu}$, affects the CMB power spectrum primarily through 
its effect on the epoch of matter-radiation equality.  There are 
substantial differences in amplitudes between those scales that 
enter the horizon during the radiation dominated era and those that 
enter the horizon later, in the matter dominated era.  Increasing 
the radiation content delays matter-radiation equality, bringing 
it closer to the epoch of recombination, suppressing the growth of 
perturbations.  As a result, the redshift of the epoch of matter-radiation 
equality, $z_{eq}$, is a fundamental observable that can be extracted 
from the CMB Power Spectrum.  $z_{eq}$ is related to the matter and 
radiation densities by,
\begin{equation}
1+z_{eq} = \rho_{\rm M}/\rho_{\rm R}.
\end{equation}
Since $\rho_{\rm R}$ depends on N$_{\nu}$, $z_{eq}$ is a function of 
both \Nnu and $\Omega_{\rm M}h^2$, leading to a degeneracy between 
these two parameters.  For a flat universe, preserving the fit to the 
CMB power spectrum when \Nnu increases, requires that $\Omega_{\rm M}$ 
and/or $H_0$ increase.  As a result of this degeneracy, the CMB power 
spectrum alone imposes only a very weak constraint on N$_{\nu}$ 
\citep{crotty03, pier03, barger03, hannestad03, ichikawa06}.  Inclusion 
of additional, independent constraints on these parameters are needed 
to break the degeneracy between \Nnu and $\Omega_{\rm M}h^2$. 

Besides affecting the epoch of matter-radiation equality, relativistic 
neutrinos leave a distinctive signature on the CMB power spectrum due 
to their free streaming at speeds exceeding the sound speed of the 
photon-baryon fluid.  This free streaming creates neutrino anisotropic 
stresses generating a phase shift of the CMB acoustic oscillations in 
both temperature and polarization.  This phase shift is unique and, for 
adiabatic initial conditions, cannot be generated by non-relativistic 
matter.  In principle, this effect can be used to break the degeneracy 
between \Nnu and $\Omega_{\rm M}h^2$, leading to tighter constraints on 
N$_{\nu}$ \citep{bash04}. 

Alternatively, since the luminosity distances of type Ia supernovae 
(SNIa) provide a constraint on a combination of $\Omega_{\rm M}$ and 
$\Omega_{\Lambda}$ complementary to that from the assumption of flatness, 
they are of value in restricting the allowed values of $\Omega_{\rm M}$.  
In concert with a bound on $H_{0}$, this, too, helps to break the 
degeneracy between \Nnu and $\Omega_{\rm M}h^{2}$.

\begin{figure}
\centerline{\epsfxsize=3truein\epsffile{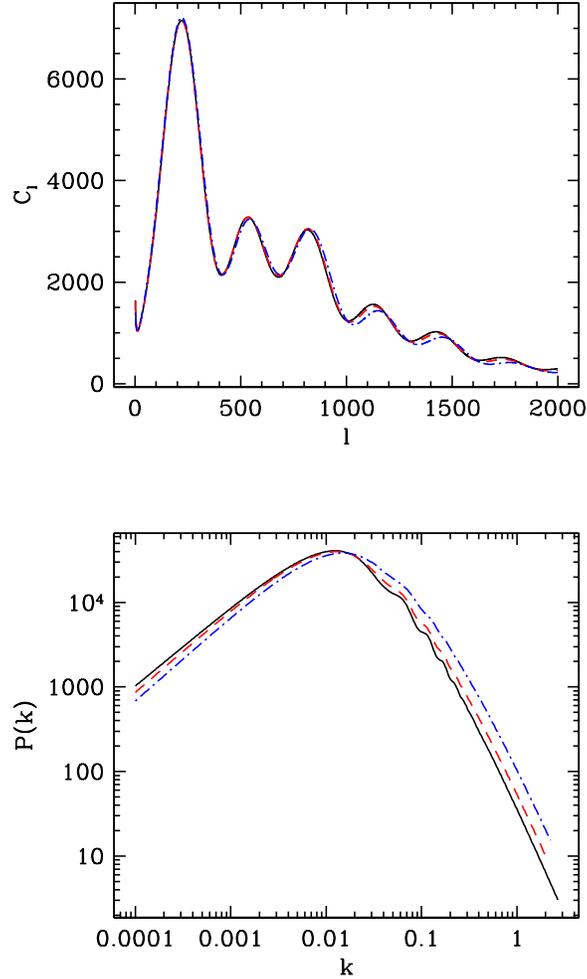}}
\caption{
The top panel shows the CMB power spectrum for the best fit models 
with \Nnu fixed at \Nnu = 1 (solid black), \Nnu = 3 (dashed red), and 
\Nnu = 5 (dot-dashed blue) illustrating its insensitivity to \Nnu in 
the absence of an independent constraint on $\Omega_{\rm M}h^{2}$. 
The bottom panel shows the matter power spectra for the same set 
of parameter values, illustrating its sensitivity to N$_{\nu}$.}
\label{fig:powerspectrum}
\end{figure}

Another way to break the degeneracy between \Nnu and $\Omega_{\rm M}h^2$ 
is to use measurements of the matter power spectrum in combination 
with the CMB power spectrum. To preserve a fit to the CMB power spectrum, 
an increase in \Nnu requires that $\Omega_{\rm M}h^2$ increase in order 
that the redshift of matter-radiation equality remain unchanged.  The 
turnover scale in the matter power spectrum is set by this connection 
between \Nnu and $\Omega_{\rm M}h^2$.  Since the baryon density is 
constrained by the CMB power spectrum, independently of N$_{\nu}$, 
increasing the radiation density (\Nnu $> 3$) requires a higher dark 
matter density in order to preserve $z_{eq}$ (in a flat universe, 
$\Omega_{\rm M} + \Omega_{\Lambda} = 1$).  Between the epoch of 
matter-radiation equality and recombination the density contrast in 
the cold dark matter grows unimpeded, while the baryon density contrast 
cannot grow.  Consequently, increasing \Nnu and $\Omega_{\rm M}h^2$, 
increases the amplitude of the matter power spectrum on scales 
smaller than the turnover scale corresponding to the size of the 
horizon at $z_{eq}$.  Data from galaxy redshift surveys can be 
used to infer the matter power spectrum, thereby constraining 
$\Omega_{\rm M}h^2$ and \Nnu.  This effect may be seen in Figure 
\ref{fig:powerspectrum} which shows that for nearly indistinguishable 
CMB power spectra, different values of \Nnu yield distinguishable 
matter power spectra.  The upper panel of Figure \ref{fig:powerspectrum} 
shows that by making suitable adjustments to the other cosmological 
parameters, specifically the matter density and the spectral index, 
CMB power spectra which are nearly degenerate up to the third peak 
of the power spectrum can be produced using very different values of 
N$_{\nu}$.  However, as the bottom panel of Figure \ref{fig:powerspectrum} 
illustrates, these models produce matter power spectra with different 
shapes, demonstrating that the matter power spectrum can be used 
to help break the degeneracy between \Nnu and $\Omega_{\rm M}h^2$.   

\subsection{Analysis And Datasets}

Before presenting our results, we describe the analysis and the 
datasets employed.

For our analysis we assume a flat, CDM cosmology with a 
cosmological constant, $\Lambda$, and three flavors of active 
neutrinos with negligible masses.  Our cosmological model is 
parametrized by seven parameters:
\be
p = \{\Omega_{\rm B}h^2, \Omega_{\rm M}h^2, h, \tau, n_S, A_S, {\rm N}_{\nu}\}.
\ee 
The contents of the universe are described by the baryon density, 
$\Omega_{\rm B}h^2$ and the matter (baryonic plus cold dark matter) 
density, $\Omega_{\rm M}h^2$.  Since a flat cosmology is assumed, 
the dark energy density and the matter density are related by 
$\Omega_{\Lambda} = 1 - \Omega_{\rm M}$.  The expansion rate of the 
universe is described by the reduced Hubble parameter, $h$ ($H_{0} 
= 100h$~kms$^{-1}$Mpc$^{-1}$).  Instantaneous reionization of the 
universe is assumed with optical depth to last scattering, $\tau$.  
$A_S$ is the amplitude of scalar perturbations and $n_S$ is the 
scalar spectral index.

The Code for Anisotropies in the Microwave Background (CAMB) 
\citep{lewis00} is used to compute the CMB power spectrum for a 
fixed set of cosmological parameters.  For a given dataset, 
our degree of belief in a set of cosmological parameters $\{p\}$ 
is quantified by the posterior probability distribution,
\be
P(p|data) \propto L(data|p)\ \Pi(p).
\ee
The likelihood $L$(data$|$p) quantifies the agreement between the 
data and the set of parameter values $\{p\}$. $\Pi$(p) represents the 
prior on cosmological parameter values before the data are considered. 

Markov Chain Monte Carlo methods are used to explore the multi-dimensional 
likelihood surface.  We use the publicly available COSMOMC code for 
our analysis \citep{lewis02}.  Flat priors are adopted for all parameters, 
along with a prior on the age of the universe, $t_0$, of $t_0 >$ 10 Gyr.

The {\it mode} of the marginalized posterior probability distribution is 
used as a point estimate and the {\it minimum credible interval} as an 
estimate of the uncertainty. The {\it minimum credible interval} selects 
the region of the parameter space around the mode, $\widehat{\theta}$, 
that contains the appropriate fraction of the volume ({\it e.g.}, 68\%, 
95\%) of the posterior probability distribution, while minimizing 
$\widehat{\theta}-\theta$.  The {\it minimum credible interval} selects 
the region of the parameter space with the highest probability 
densities\footnote{The commonly used GETDIST analysis package uses the 
mean of the marginalized posterior probability distribution as a point 
estimate and gives uncertainty estimates based on the {\it central 
credible interval} \citep{hamann07}. These estimates are identical 
for Gaussian probability distributions but differ significantly for 
non-Gaussian distributions, particularly for asymmetric probability 
distributions.}. 

Our primary dataset is the CMB data from the Wilkinson Microwave 
Anisotropy Probe (WMAP) accumulated from five years of observations 
\citep{hinshaw08,nolta08}.  For the WMAP data, likelihoods are computed using 
the code provided on the LAMBDA webpage\footnote{http://lambda.gsfc.nasa.gov}.  
There are a number of ground and balloon based CMB experiments whose 
high angular resolution probe smaller scales than those probed by WMAP.  
These are more sensitive to the higher order acoustic oscillations 
beyond the third peak in the CMB power spectrum.  In particular, the 
2008 results from the Arcminute Cosmology Bolometer Array Receiver 
(ACBAR) \citep{reichardt08} impose the strongest constraints at present 
on the CMB power spectrum at small angular scales.  In our analysis, 
data from the following CMB experiments are used: BOOMERANG 
\citep{piacentini06,montroy06}, ACBAR \citep{reichardt08}, CBI 
\citep{readhead04,mason03}, VSA \citep{dickenson04}, 
MAXIMA \citep{hanany00} and DASI \citep{halverson01}.

To help break the degeneracy between \Nnu and $\Omega_{\rm M}h^2$, 
we adopt a Gaussian prior on the Hubble parameter, $H_0 = 72 \pm 
8$~kms$^{-1}$Mpc$^{-1}$ from the Hubble Space Telescope Key Project 
\citep{freedman01}.

Since the luminosity distances of type Ia supernovae (SNIa) provide 
a constraint on a combination of $\Omega_{\rm M}$ and $\Omega_{\Lambda}$, 
they are of value in restricting the allowed values of $\Omega_{\rm M}$,  
helping to break the degeneracy between \Nnu and $\Omega_{\rm M}h^{2}$. 
We use the luminosity distances for 115 type Ia supernovae measured 
by the Supernova Legacy Survey (SNLS) \citep{astier06}.  For each 
supernova, the observed luminosity distance is compared to that 
predicted for a given set of cosmological parameters.

We use the matter power spectrum inferred from galaxy redshift 
surveys such as the Sloan Digital Sky Survey (SDSS) \citep{tegmark04, 
tegmark06} and the 2dFGRS \citep{cole05}.  In using the matter power 
spectrum to infer cosmological parameters, it has become clear 
that the constraint on $\Omega_{\rm M}$ is sensitive to the choice 
of length scales on which the power spectrum is measured.  The power 
spectrum on smaller scales prefers higher values of $\Omega_{\rm M}$, 
provided that those scales are correctly described by linear 
perturbation growth and scale independent galaxy bias 
\citep{cole05,percival07}.  For example, using the SDSS LRG power 
spectrum \citep{tegmark06} in combination with the 3 year WMAP 
dataset \citep{sp07}, \cite{dunkley08} find a disagreement between 
the matter density inferred on scales with $k\le0.1h$~Mpc$^{-1}$ 
and $k\le0.2h$~Mpc$^{-1}$.  \cite{hamann07} have shown that the 
constraint on \Nnu obtained from the matter power spectrum is 
affected similarly, resulting from the correlation between 
$\Omega_{\rm M}$ and N$_{\nu}$.  Therefore, here we only use 
matter power spectrum data on scales that are likely to be 
safely linear.  We truncate the matter power spectrum at 
$k=0.07h$~Mpc$^{-1}$, keeping data only on scales with 
$k\le0.07h$~Mpc$^{-1}$.  We assume that galaxy bias is 
constant and scale independent for these scales.

\subsection{\Nnu And The Lyman-$\alpha$ Forest}
\label{lya}

Measurements of the flux power spectrum of the Lyman-$\alpha$ 
forest in QSO absorption spectra can be used to reconstruct the 
matter power spectrum on small scales \citep{croft98,mcdonald00}. 
Observations of the SDSS Ly-$\alpha$ flux power spectrum have 
been used to constrain the linear matter power spectrum at $z\sim3$ 
\citep{mcdonald05}.  When these constraints are combined with CMB 
power spectrum data, they favor considerably higher values of 
\Nnu than those we find here (\S5).  For example, \cite{seljak06} 
found N$_{\nu}=5.2$ and a 95\% range $3.4 \leq$ N$_{\nu} \leq 7.3$ 
and \cite{hamann07} found N$_{\nu}=6.4$ and a 95\% range $3.2 \leq$ 
N$_{\nu} \leq 11$, the difference being accounted for by their 
different combinations of data sets.  Each of these excludes 
the standard model value of \Nnu = 3 at more than 95\% confidence.

As discussed earlier and shown in the lower panel of Figure 
\ref{fig:powerspectrum}, the principal effect of an increase in 
\Nnu is to increase the amplitude of the matter power spectrum 
on scales smaller than those corresponding to the horizon at 
matter-radiation equality, $z_{eq}$.  The $\Lambda$CDM fits to 
the Lyman-$\alpha$ forest data prefer higher amplitudes of density 
fluctuations on small scales compared to those expected from 
measurements of the WMAP power spectrum \citep{viel06}, favoring 
higher values of N$_{\nu}$.  This effect is seen in Figure 
\ref{fig:powerspectrum} which shows that for nearly indistinguishable 
CMB power spectra, different values of \Nnu yield distinguishable
matter power spectra.  To preserve $z_{eq}$ and the fit to the CMB, 
the model with \Nnu = 1 has a lower value of $\Omega_{\rm M}h^2$ and, 
therefore, the matter power spectrum for that model has a lower 
amplitude on scales smaller than the scale corresponding to the 
horizon at the epoch of matter-radiation equality.  Conversely, 
the model with \Nnu = 5 has a higher value of $\Omega_{\rm M}h^2$ 
and the matter power spectrum for that model has a higher amplitude 
on scales smaller than the scale corresponding to the horizon at 
$z_{eq}$.

Before reaching any conclusions about \Nnu based on the Lyman-$\alpha$ 
forest data, it is worth noting that assumptions about the thermal 
state of the IGM play an important role in reconstructing the matter 
power spectrum from the Lyman-$\alpha$ forest flux power spectrum.  
\cite{bolton07} compare measurements of the Lyman-$\alpha$ forest flux 
probability distribution by \cite{kim07} to hydrodynamic simulations 
of the Lyman-$\alpha$ forest, finding evidence for an {\it inverted} 
temperature-density relation for the low density intergalactic medium.  
\cite{bolton07} suggest that \heii reionization could be a possible 
physical mechanism for achieving an inverted temperature-density 
relation.  Such an inversion would result in a smaller amplitude 
of the matter power spectrum for a given observed flux power 
spectrum, thereby alleviating the tension with the other data 
sets which tended to drive \Nnu to high values.  However, in their 
power spectrum fits \cite{mcdonald05} marginalize over equation of 
state parameters, so this explanation of the tension may not be 
entirely satisfactory.  Future studies of the Lyman-$\alpha$ forest 
may weaken or strengthen the evidence for a discrepancy.  For 
these reasons, we do not use data from the Lyman-$\alpha$ forest 
in our analysis.

\subsection{CMB And LSS Constraints On N$_{\nu}$ And $\eta_{\rm B}$}

\begin{figure}
\centerline{\epsfxsize=3truein\epsffile{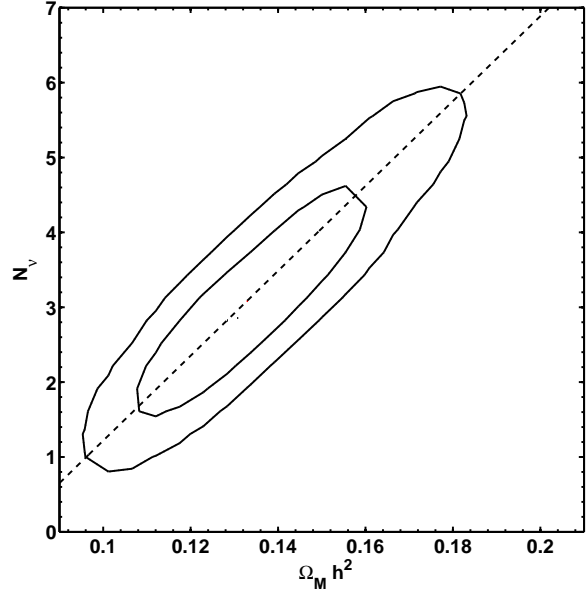}}
\caption{
The 68\% and 95\% contours in the \Nnu - $\Omega_{\rm M}h^2$ plane
inferred from the combination of the  WMAP-5yr data, small scale 
CMB data, luminosity distances of SNIa and the HST Key Project prior 
on $H_0$. The dashed line shows the locus of points corresponding 
to the same value of $z_{eq} (= 3144$), illustrating the degeneracy 
between these two parameters.  As the contours reveal, this 
degeneracy may be broken if complementary data is used to constrain 
$\Omega_{\rm M}h^2$.}
\label{fig:omh2nnu}
\end{figure}

According to \cite{dunkley08} the WMAP 5 year data only impose a 
lower limit on \Nnu of \Nnu $> 2.3$ but, according to \cite{komatsu08}, 
do not lead to an upper limit due to the degeneracy between \Nnu and 
$\Omega_{\rm M}h^2$.  Inclusion of data from small scale CMB experiments 
do not break this degeneracy.  Figure \ref{fig:omh2nnu} illustrates this 
degeneracy, as well as how it may be broken by non-CMB constraints on 
$\Omega_{\rm M}h^2$.  The dashed line in Fig.~\ref{fig:omh2nnu} is the 
locus of points with constant $z_{eq} = 3144$.  Figure \ref{fig:omh2nnu} 
shows the joint probability distributions of \Nnu and $\Omega_{\rm M}h^2$ 
inferred from the WMAP 5yr data and small scale CMB experiments, 
supplemented by independent data from measurements of SNIa luminosity 
distances and the HST Key Project prior on $H_0$ are used to bound 
$\Omega_{\rm M}h^2$. The range of \Nnu is now limited.

\begin{figure}
\centerline{\epsfxsize=2.5truein\epsffile{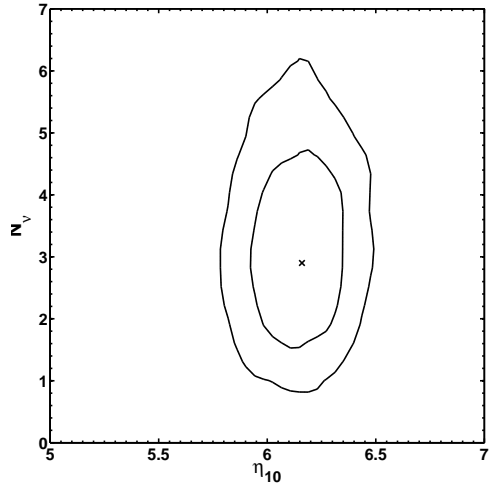}}
\caption{
The 68\% and 95\% contours in the N$_{\nu} - \eta_{10}$ plane inferred 
from the combination of the WMAP-5yr data, small scale CMB data, SNIa 
luminosity distances, and the HST Key Project prior on $H_0$ (see the 
text and Table 1).} 
\label{fig:nnue10cmb}
\end{figure}

\begin{figure}
\centerline{\epsfxsize=2.5truein\epsffile{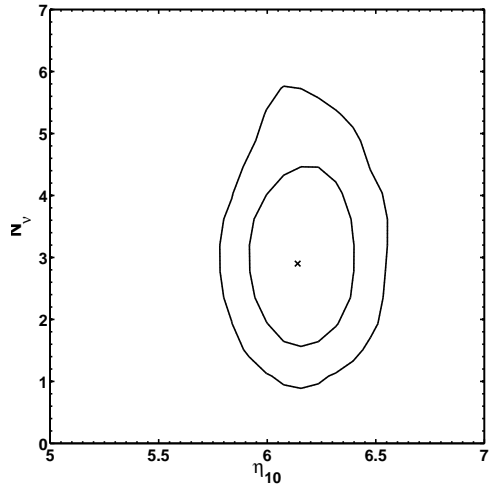}}
\caption{
The 68\% and 95\% contours in the N$_{\nu} - \eta_{10}$ plane derived 
using the WMAP 5-year data, small scale CMB data, SNIa and the HST 
Key Project prior on $H_0$ along with matter power spectrum data 
from 2dFGRS and SDSS LRG(see the text and Table 1).}
\label{fig:nnue10cmblss}
\end{figure}

Our constraints on \Nnu and $\eta_{10}$ from various CMB and 
LSS datasets and combinations of them are summarized in Table 1
 and in Figures 5 -- 7.  

Using the WMAP 5 year data in combination with data from other CMB 
experiments (ACBAR, BOOMERANG, CBI, DASI, MAXIMA and VSA), along with 
the HST Key Project prior on $H_0$ and luminosity distance measurements 
of type Ia supernovae, we find central values \Nnu = $2.9$ and \eten = 
$6.2$, along with 68\% (95\%) ranges of $2.0<$~\Nnu$ <4.1$ ($1.3<$~\Nnu$<5.4$) 
and $6.0<$~\eten$<6.3$ ($5.9<$~\eten$<6.4$).  Figure \ref{fig:nnue10cmb} shows 
the joint probability distribution of \Nnu and \eten for this combination 
of CMB datasets.

Adding the 2dFGRS power spectrum \citep{cole05} to the CMB 
power spectrum from the WMAP 5 year dataset, along with 
ground based CMB experiments (see above), the HST prior 
on $H_0$ and, luminosity distance measurements from SNIa, 
we find central values \Nnu = $3.0$ and \eten = $6.1$, along 
with 68\% (95\%) ranges $2.1<$~\Nnu$<4.2$ ($1.3<$~\Nnu$<5.2$) and 
$6.0<$~\eten$<6.3$ ($5.9<$~\eten$<6.4$) respectively; see Table 1.  
Replacing the 2dFGRS power spectrum with the SDSS LRG power 
spectrum \citep{tegmark06}, we obtain very similar results: 
\Nnu = $2.8$, $2.1<$~\Nnu$<3.9$ ($1.5<$~\Nnu$<5.2$) and, \eten = $6.2$, 
$6.1<$~\eten$<6.3$ ($5.9<$~\eten$<6.5$).  In contrast, the SDSS DR2 
\citep{tegmark04} power spectrum favours considerably higher 
values of \Nnu compared to those inferred from the 2dFGRS 
and the SDSS (LRG) power spectra.

Our best constraints on \Nnu and $\eta_{10}$ from the CMB and LSS data 
are derived by combining the following datasets - WMAP 5-year data, 
ground and balloon based CMB experiments (BOOMERANG, ACBAR, CBI, 
VSA, MAXIMA and DASI), the HST prior on $H_0$, the SNIa luminosity 
distance measurements and, the matter power spectrum on large scales 
inferred from the 2dFGRS and SDSS (LRG) data.  Constraints on the 
matter power spectrum from the Lyman-$\alpha$ forest are not included 
for the reasons discussed above in \S\ref{lya}.  Using these datasets, 
we obtain (see Table 1):
\be
\eta_{10}=6.1^{+0.2 +0.3}_{-0.1 -0.2}.
\ee
\be
{\rm N}_{\nu}=2.9^{+1.0 +2.0}_{-0.8 -1.4}.
\ee
At present the combined CMB and LSS data provide the best baryometer,
determining the baryon density to better than 3\%, but only a relatively 
weak chronometer, still allowing a large range in $S$ ($0.87 \leq S \leq 
1.14$ at 95\% confidence).  Within their uncertainties, the CMB/LSS 
data, which probe the Universe at $\ga 400,000$ years, are consistent 
with BBN, which provides a window on the Universe at $\la 20$ minutes.

\begin{deluxetable*}{l c c}
\tablecolumns{3} 
\large
\tablecaption{N$_{\nu}$ and $\eta_{10}$ from different datasets}
\tablehead{\colhead{Dataset} & \colhead{N$_{\nu}$} & \colhead{$\eta_{10}$}}
SPMquot15pt
\startdata

\smallskip
BBN (\textrm{Y$_{\rm P}$} \& \textrm{$y_{\rm DP}$}) &$2.4^{+0.4 +0.9}_{-0.4-0.8}$ &$5.7^{+0.4 +0.8}_{-0.4 -0.9}$\\
\smallskip
WMAP(1yr)+HST &$2.8^{+4.5 +5.5}_{-1.8 -1.9}$ &$6.1^{+0.5 +1.8}_{-0.3 -0.7}$\\
\smallskip
WMAP(3yr)+HST+SN &$2.9^{+2.1 +4.0}_{-1.2 -2.1}$ &$6.06^{+0.23 +0.42}_{-0.20 -0.39}$ \\
\smallskip
WMAP(3yr)+CMB(07)+HST+SN &$2.5^{+1.7 +3.8}_{-1.2 -1.7}$ &$6.13^{+0.21 +0.39}_{-0.18 -0.36}$\\
\smallskip
2dFGRS+WMAP(3yr)+CMB(07)+HST+SN &$2.9^{+1.4 +2.3}_{-1.2 -2.3}$ &$6.11^{+0.20 +0.38}_{-0.15 -0.32}$\\
\smallskip
SDSS(DR2)+WMAP(3yr)+CMB(07)+HST+SN &$3.7^{+1.6 +3.3}_{-1.4 -2.4}$ &$6.15^{+0.15 +0.35}_{-0.20 -0.39}$\\
\smallskip
SDSS(LRG)+WMAP(3yr)+CMB(07)+HST+SN &$2.0^{+1.2 +2.6}_{-1.2 -1.8}$ &$6.12^{+0.19 +0.36}_{-0.16 -0.34}$\\
\smallskip
WMAP(5yr)+HST+SN &$3.9^{+2.0 +4.5}_{-1.2 -2.4}$ &$6.11^{+0.16 +0.33}_{-0.16 -0.33}$ \\
\smallskip
WMAP(5yr)+CMB+HST+SN &$2.9^{+1.2 +2.5}_{-0.9 -1.6}$ &$6.16^{+0.14 +0.25}_{-0.16 -0.30}$\\
\smallskip
2dFGRS+WMAP(5yr)+CMB+HST+SN &$3.0^{+1.2 +2.2}_{-0.9 -1.7}$ &$6.14^{+0.16 +0.27}_{-0.14 -0.25}$\\
\smallskip
SDSS(LRG)+WMAP(5yr)+CMB+HST+SN &$2.8^{+1.1 +2.4}_{-0.7 -1.3}$ &$6.16^{+0.14 +0.27}_{-0.14 -0.27}$\\
\smallskip
SDSS(LRG)+2dFGRS+WMAP(5yr)+CMB+HST+SN &$2.9^{+1.0 +2.0}_{-0.8 -1.4}$ &$6.14^{+0.16 +0.30}_{-0.11 -0.25}$\\
\smallskip
BBN+SDSS(LRG)+2dFGRS+WMAP(5yr)+CMB+HST+SN &$2.5^{+0.4 +0.7}_{-0.4 -0.7}$ &$6.11^{+0.12 +0.26}_{-0.13 -0.27}$\\ 
\smallskip
\enddata
\tablecomments{Best fits and 68\% and 95\% confidence intervals for \Nnu and 
\eten from our principal datasets.  WMAP refers to the CMB 
power spectrum data from the WMAP experiment and CMB to the
data from ACBAR+BOOM+CBI+VSA+MAXIMA+DASI. CMB(07) uses the 
2007 ACBAR dataset \citep{kuo07} while CMB uses the 2008 
ACBAR dataset \citep{reichardt08}. HST refers to the prior 
on $H_0$ from the HST Key Project.  SN stands for the SNIa
luminosity distance measurements.  SDSS (LRG) and 2dF refer 
to the respective LSS matter power spectra, truncated at 
$k=0.07h$~Mpc$^{-1}$.}

\end{deluxetable*}

\section{Comparing The BBN And CMB/LSS Constraints}

\begin{figure*}
\centerline{\epsfxsize=5truein\epsffile{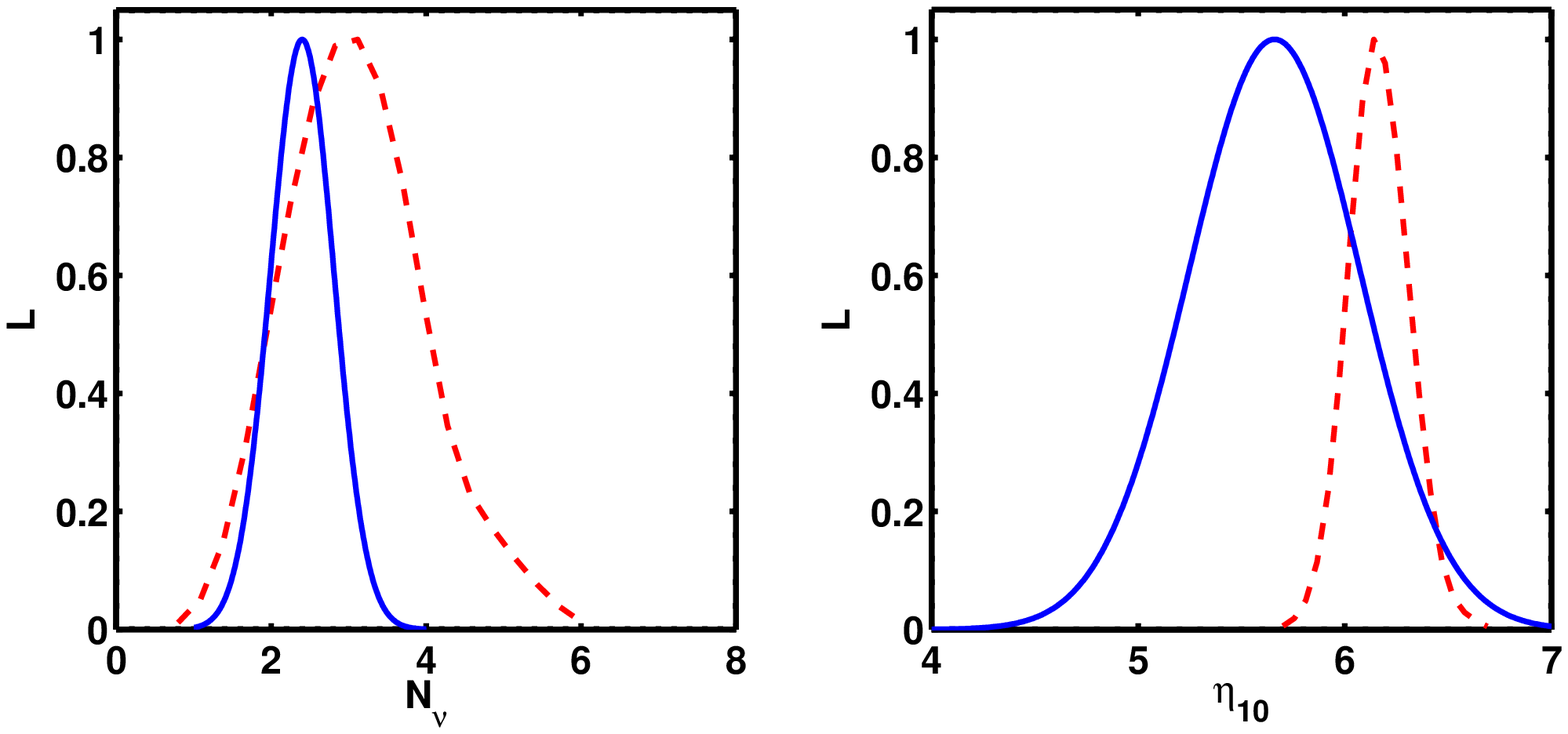}}
\caption{}
On the left (red, dashed), the probability distribution of \Nnu 
inferred from the combination of the WMAP 5-year data, small scale CMB 
data, SNIa and the HST Key Project prior on $H_0$ and matter power 
spectrum data from 2dFGRS and SDSS LRG.  The solid blue curve is
the BBN  (D plus \4he) distribution.  On the right (same line types 
and colors),  are the probability distributions of $\eta_{10}$ using 
the same data sets.
\label{fig:2panel}
\end{figure*}

\begin{figure*}
\centerline{\epsfxsize=6truein\epsffile{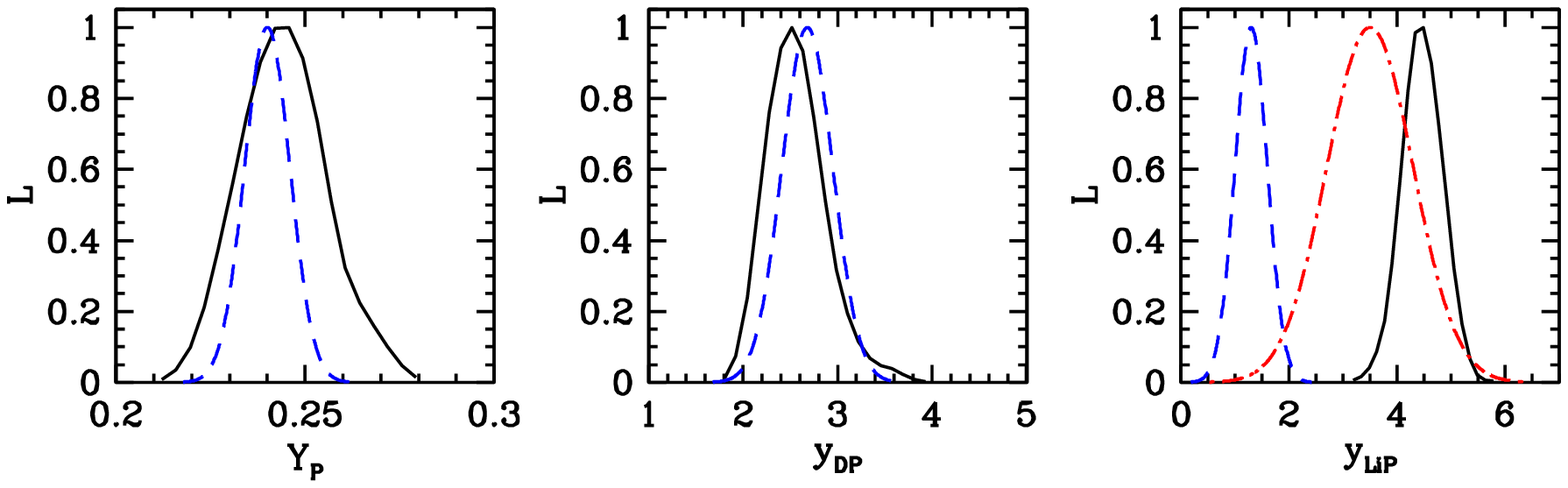}}
\caption{}
The solid black curves show the probability distributions for the 
primordial abundances of D, \4he and \7li derived from the values 
of $\eta_{10}$ and N$_{\nu}$ inferred from the CMB, LSS, SNIa and 
the HST prior and matter power spectrum data from 2dFGRS and SDSS 
LRG.  The blue dashed curves show the probability distributions
for the observationally-inferred primordial abundances of D, \4he 
and \7li; see \S3.1. The red, dot-dashed curve in the far right panel 
is the \7li abundance from \cite{korn06}. 
\label{fig:bbncomp}
\end{figure*}

Using the ranges of $\eta_{10}$ and N$_{\nu}$ allowed by the CMB and 
LSS data, the BBN-predicted primordial abundances of \4he, D and \7li 
may be inferred.  Figure \ref{fig:bbncomp} compares these constraints 
to the observationally-inferred primordial abundances adopted in \S3.1.  
The CMB/LSS-inferred BBN abundances of D and \4he are in excellent 
agreement, within the errors, with the observationally-inferred relic 
abundances.  For the central values of $\eta_{10} = 6.14$ and \Nnu = 
2.9, the BBN-predicted deuterium abundance of $y_{\rm D\rm P} =2.54$ 
is, within the errors, in agreement with its observationally-inferred 
primordial value of $y_{\rm D\rm P} = 2.68$.  For \4he, the BBN-predicted 
mass fraction is \Yp = 0.247, slightly high compared to the central 
value of the primordial abundance adopted in \S3.1, \Yp = 0.240, but 
within 1.2$\sigma$ of it and, completely consistent with the 
evolution-model independent upper bound presented in eq.~15, 
Y$_{\rm P} < 0.251\pm0.002$.   

For \7li the BBN-predicted best fit from the CMB/LSS data is 
[Li]$_{\rm P}$ = 2.66, considerably higher than the value 
([Li]$_{\rm P} = 2.1\pm 0.1$) determined from observations of 
metal-poor halo stars (\citet{ryan}, \citet{asplund06}) without any 
correction for depletion, destruction, or gravitational settling. 
If, however, the correction proposed by \cite{korn06} is applied, 
the predicted and observed \7li abundances may, perhaps, be reconciled, 
as may be seen from the lower panel of Fig.~\ref{fig:bbncomp}. 
It remains an open question whether this lithium problem is 
best resolved by a better understanding of stellar physics or, 
if it is providing a hint of new physics beyond the standard model.

\section{Conclusions}

\begin{figure*}
\centerline{\epsfxsize=5truein\epsffile{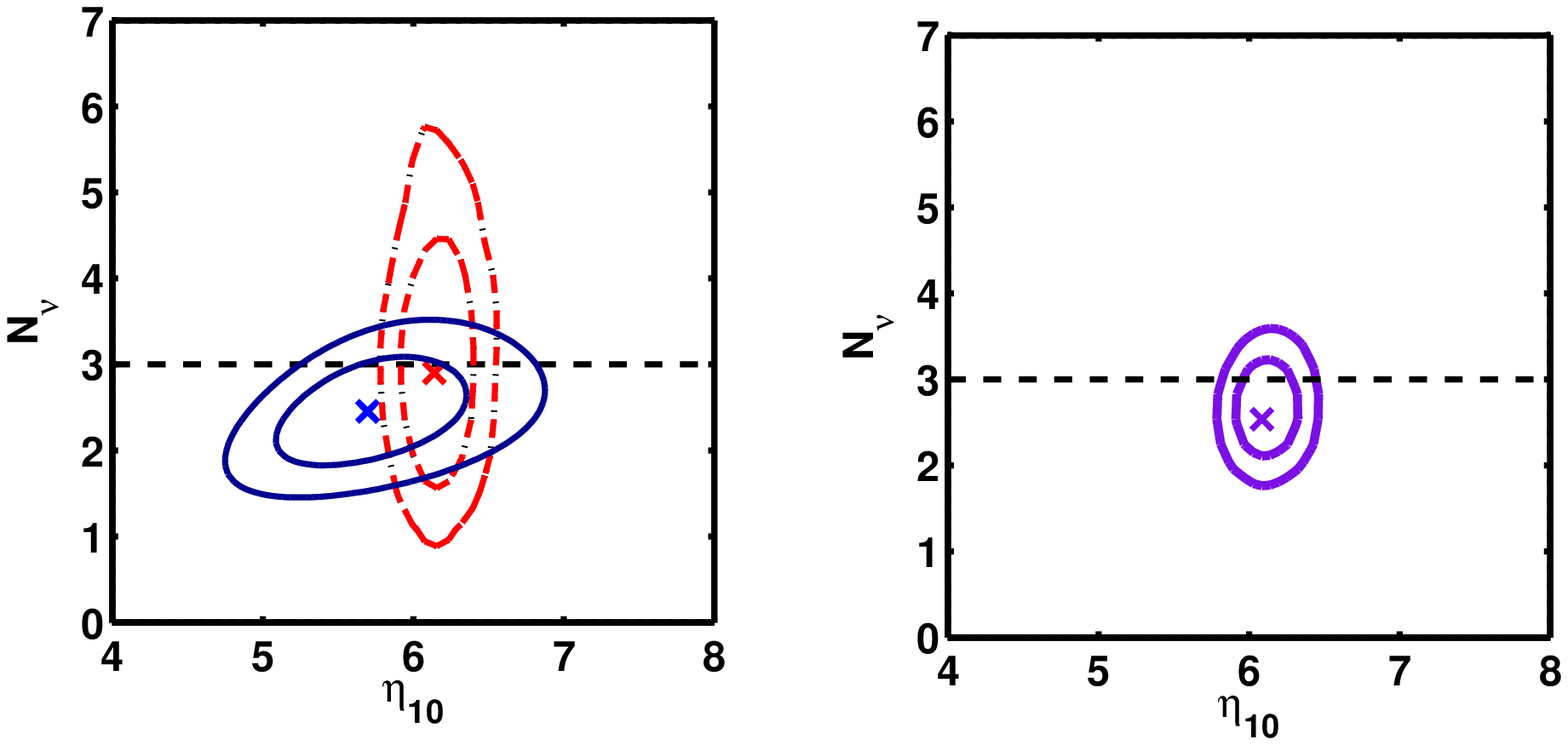}}
\caption{}
(Left) In blue (solid), the 68\% and 95\% contours in the \Nnu - $\eta_{10}$ 
plane derived from a comparison of the observationally-inferred and 
BBN-predicted primordial abundances of D and \4he (see Figure 2).  
In red (dashed), the 68\% and 95\% contours derived from the combined 
WMAP 5-year data, small scale CMB data, SNIa, and the HST Key Project 
prior on $H_0$ along with matter power spectrum data from 2dFGRS and 
SDSS LRG (see the text and Table 1). (Right) The 68\% and 95\% joint 
BBN-CMB-LSS contours in the N$_{\nu} - \eta_{10}$ plane.
\label{fig:nnue10}
\end{figure*}

While our CMB+LSS constraints on $\eta_{\rm{B}}$ and \Nnu are consistent 
with most previous analyses \citep{crotty03,pier03,hannestad03,barger03,seljak06,ichikawa06,sp07,mangano07,hamann07,dunkley08,komatsu08},
they are tighter because we have used more 
and/or more recent data.  However, as discussed above in \S\ref{lya}, 
analyses that included the Lyman-$\alpha$ forest data generally find 
higher values of N$_{\nu}$.

Until the advent of WMAP and the other ground- and balloon-based CMB
experiments, BBN provided the best baryometer (mainly from deuterium)
and chronometer (mainly from helium-4).  As may be seen from Table 1,
while the WMAP first year data provided a competitive baryometer, it
offered a relatively poor chronometer.  This improves with the WMAP
3 and 5 year data, especially when they are combined with the other 
CMB and LSS data.  These now lead to a determination of the baryon 
density at the $\sim 2-3 \%$ level, a factor of $\ga 2$ better than 
that from BBN.  However, although the CMB/LSS constraint on \Nnu has 
improved significantly and, it is consistent with that from BBN, it 
remains weaker than the BBN constraint by a factor of $\sim 2$.  For
some time now BBN has clearly established at high confidence that 
\Nnu $> 1$ when the Universe was $\sim 20$ minutes old.  For example, 
using a slightly different estimate of the primordial helium mass 
fraction, \cite{barger03} found \Nnu $> 1.7$ at 95\% confidence (see 
also \cite{ijmpe}).  The more recent WMAP 5 year data, combined with 
the recent ACBAR and other CMB and LSS datasets, the CMB/LSS now 
confirm that \Nnu $> 1$ (or, \Nnu $> 2$ \citep{dunkley08}) when the 
Universe was $\ga 400$~kyr old.

As may be seen from Table 1 and from Figures \ref{fig:2panel}, \ref{fig:bbncomp}, 
and from the left-hand panel of Figure \ref{fig:nnue10}, BBN and the CMB, 
which probe physics at widely separated epochs in the evolution of the 
Universe, are in excellent agreement.  This permits constraints on any
{\it differences} in physics between BBN and recombination and/or the
present epoch.  For example, since baryons are conserved, $\eta_{\rm B}$ 
relates the number of thermalized black body photons in a comoving volume 
at different epochs (see figure \ref{fig:2panel}), constraining any post-BBN 
entropy production.  Our results from the CMB/LSS and from BBN imply,
\be
{N_{\gamma}^{\rm CMB} \over N_{\gamma}^{\rm BBN}} = 0.92 \pm 0.07.
\ee 
This ratio is consistent with 1 at $\sim$1$\sigma$ and places an interesting 
upper bound on any post-BBN entropy production.

Alternatively, late decaying particles could produce relativistic 
particles (radiation), but not necessarily thermalized black body 
photons (see, for example, \cite{ichikawa07}).  Deviations from the 
standard model radiation density can be parametrized by the ratio 
of the radiation density, $\rho'_{\rm R}$ to the standard model 
radiation density, $\rho_{\rm R}$.  In the post-\epm annihilation 
universe (see eq.~9),
\be
R = S^2 = {\rho'_{\rm R} \over \rho_{\rm R}} = 1 + 0.134\Delta{\rm N}_{\nu}.
\ee
Comparing this ratio, at BBN and at recombination (see figure \ref{fig:2panel}), 
any post-BBN production of relativistic particles can be constrained.
\be
{R_{\rm CMB} \over R_{\rm BBN}} = 1.07^{+0.16}_{-0.13}
\ee
This ratio, too, is consistent with 1 within 1$\sigma$, placing an 
upper bound on post-BBN production of relativistic particles.

Although the non-standard expansion rate has been parametrized in terms
of an equivalent number of additional species of neutrinos, we have 
emphasized that a non-standard expansion rate need not be related to 
extra (or fewer) neutrinos.  For example, deviations from the standard 
expansion rate could occur if the value of the early-Universe gravitational 
constant, G$_{\rm N}$ were different in the from its present, locally-measured 
value \citep{yang79, boes85, accetta90, cyburt05}.  For the standard radiation 
density with three species of light, active neutrinos, the constraint on 
the expansion rate can be used to constrain variations in the gravitational 
constant, G$_{\rm N}$.  From BBN, 
\be
S^{2} = {G^{\rm BBN}_{\rm N} \over G_{\rm N}} = 0.91 \pm 0.07
\ee 
and at the epoch of the recombination,
\be
S^{2} = {G^{\rm CMB}_{\rm N} \over G_{\rm N}} = 0.99^{+0.13}_{-0.11},
\ee
consistent with no variation in G at the $\sim$1$\sigma$ level.

The agreement between $\eta_{\rm B}$ and \Nnu evaluated from BBN
($\sim$~20 minutes) and from the CMB/LSS ($\ga$~400 kyr) is shown
in the left hand panel of Figure \ref{fig:nnue10}. As Figure 
\ref{fig:nnue10} illustrates, BBN and the CMB/LSS, which probe 
the Universe at widely separated epochs in its evolution, are 
completely consistent.  As already noted, at present the CMB 
is a better baryometer while BBN remains a better chronometer.  
Since these independent constraints from the CMB/LSS and BBN are 
in very good agreement, we may combine them to obtain the joint 
fit in Table 1 and shown in the right hand panel of Figure 
\ref{fig:nnue10}.  We note that while the best-fit value of \Nnu 
is less than $3$, this is not statistically significant since 
the results are consistent with the standard model of three 
species of active neutrinos at the $\sim$1$\sigma$ level.

Of course, our BBN results are sensitive to the relic 
abundances we have adopted.  For comparison, we have repeated our 
analysis for an alternate set of primordial abundances.  For deuterium, 
we adopted the \cite{omeara06} results based on the weighted mean 
of the log($y_{D}$) values, $y_{D} = 2.84^{+0.27}_{-0.25}$ and, 
for \4he we chose, somewhat arbitrarily, \Yp = $0.247\pm0.004$.  
For this alternate abundance set the BBN-predicted value of 
$\eta_{\rm B}$ is virtually unchanged from our previous result, 
$\eta_{10} = 5.7\pm0.3$, while \Nnu = $2.9\pm0.3$ is much closer to 
the standard model expectation.  As a result, while the constraint 
on entropy production (eq.~26) is unchanged, $R_{\rm CMB}/R_{\rm 
BBN} = 1.00^{+0.16}_{-0.13}$ and G$^{\rm BBN}_{\rm N}/$G$_{\rm N} 
= 0.99^{+0.13}_{-0.11}$.  When the alternate BBN constraints are 
combined with those from the CMB and LSS, we find $\eta_{10} = 
6.09^{+0.12}_{-0.13}$ and \Nnu = $2.9\pm0.3$.

Very recent observations of deuterium in the high redshift, low metallicity Damped Lyman-$\alpha$ absorber
by \cite{pettini08} lead to a deuterium abundance very close to the mean of the previous six abundances used 
in this paper. As a result, the change in $y_{D}$ is very small ($y_{D}= 2.70$ rather than $y_{D}=2.68$ adopted in this paper). Consequently, the change 
in the parameters inferred from it (N$_{\nu}$ and $\eta_{10}$) is also very 
small. 

Future CMB experiments will improve the constraint on N$_{\nu}$ 
by measuring the neutrino anisotropic stress more accurately. 
According to \cite{bash04}, PLANCK should determine N$_{\nu}$ to 
an accuracy of $\sigma(N_{\nu}) \sim 0.24$ and CMBPOL, a satellite 
based polarization experiment, might improve it further to 
$\sigma(N_{\nu}) \sim 0.09$, independent of the BBN constraints.  
In this context, we note that such tight constraints on \Nnu will 
be sensitive to the value of the \4he abundance adopted in the 
CMB analysis.  To achieve these projected accuracies, it will 
no longer be sufficient to fix \Yp in advance.  Rather, \Yp should 
be solved for in concert with the other cosmological parameters.  
To this end, we point out that for $\eta_{10} \approx 6$, \Nnu 
$\approx 3$ (suggested by our results), a very good, simple 
approximation to \Yp is provided by eq.~11 \citep{kneller04, 
steigman07}.

\begin{center}
ACKNOWLEDGMENTS
\end{center}
This research is supported at The Ohio State University by a grant 
(DE-FG02-91ER40690) from the US Department of Energy.  We thank D. 
Weinberg for a careful reading of the manuscript and for suggestions 
which led to an improved manuscript.  We also thank R. Cyburt, S. Dong, 
J. Dunkley, J. Hamann, S. Hannestad, A. Lewis, P. McDonald, M. Pettini, G. Raffelt, 
and Y. Wong for useful discussions.  We thank the Ohio Supercomputer 
Center for the use of a Cluster Ohio Beowulf cluster in this research.  

\singlespace

\clearpage

\end{document}